\documentclass[fleqn,10pt]{wlscirep}
\usepackage[utf8]{inputenc}
\usepackage[T1]{fontenc}
\title{Glycocalyx Cleavage Boosts Erythrocytes Aggregation}

\author[1,2,*]{Mehdi Abbasi}
\author[1,*]{Min Jin}
\author[3]{Yazdan Rashidi}
\author[1]{Lionel Bureau}
\author[1]{Daria Tsvirkun}
\author[1,"]{Chaouqi Misbah}

\affil[1]{Univ. Grenoble Alpes, CNRS, LIPhy, F-38000 Grenoble, France}
\affil[2]{Aix Marseille Université, CNRS, Centre Interdisciplinaire de Nanoscience de Marseille, Turing Centre for Living Systems, Marseille 13009, France}
\affil[3]{Dynamics of Fluids, Department of Experimental Physics, Saarland University, 66123 Saarbrucken, Germany}

\affil["]{chaouqi.misbah@univ-grenoble-alpes.fr}
\affil[*]{Equally contributed to this work}

\begin{abstract}
The glycocalyx is a complex layer of carbohydrate and protein molecules that surrounds the cell membrane of many types of mammalian cells. It serves several important functions, including cell adhesion and communication, and maintain cell shape and stability, especially in the case of erythrocytes. Alteration of glycocalyx composition represents a cardiovascular health threatening. For example, in diabetes mellitus glycocalyx of erythrocytes and  of endothelial cells is known to be impaired, a potential source of blood occlusion in microcirculation, which may  lead to blindness,  and renal failure of patients. The impact of glycocalyx impairment on erythrocyte aggregation remains a largely unexplored research area. We conduct here  in vitro-experiments in microfluidic devices in order to investigate erythrocytes aggregation  incubated with amylase, an enzyme that partially breaks down glycocalyx molecules. It is found that incubation of erythrocytes by amylase  leads to  a   dramatic increase of their aggregation and  stability and alters the  aggregates morphologies. Confocal microscopy analysis reveals a significant degradation of the glycocalyx layer, correlated with enhanced erythrocytes aggregation. An increased erythrocyte aggregation in vivo should affect oxygen and other metabolites delivery to organs and tissues. This study brings new elements about elucidation of microscopic origins of erythrocyte aggregation and their potential impact on cardiovascular pathologies. 
\end{abstract}
\begin{document}

\flushbottom
\maketitle
%
%
\thispagestyle{empty}


\section{Introduction}

Human red blood cells (RBCs) exhibit the intriguing ability to form  aggregates in the presence of various plasma proteins (e.g., fibrinogen and immunoglobulins) and polymers such as Dextran in vitro \cite{baskurt2011red,bosek,bertoluzzo1999kinetic,brust2014plasma}. This phenomenon has garnered substantial attention through numerical simulations \cite{ju2013effect,bagchi2005computational,T.Wang,flormann2017buckling,PhysRevFluids.4.103601,abbasi2021erythrocyte}. Under normal physiological conditions, RBC aggregation is reversible and plays a crucial role in microcirculation regulation, with blood viscosity being closely linked to the level of RBC aggregation. However, in pathological conditions such as hematological diseases, cardiovascular diseases, and metabolic disorders like diabetes mellitus, persistent RBC aggregation may occur, potentially leading to blood vessel occlusion and severe health complications \cite{groeneveld1999relationship, prasad1993oxygen,rogers1992decrease}.

Two proposed models \cite{baskurt2011red} aim to explain RBC aggregation mechanisms: I) the depletion layer model, suggesting aggregation arises due to the depletion layer of macromolecules, resulting in entropic attraction; II) the bridging effect model, proposing the formation of bonds between RBCs through intermediate molecules like fibrinogen. Nevertheless, a comprehensive understanding of how these elements intricately contribute to firm RBC aggregation remains elusive.

The glycocalyx, a multifunctional sugar layer covering the surface of all mammalian cells, is primarily composed of glycoproteins, glycolipids, and proteoglycans \cite{Glyco}. While most research on glycocalyx has focused on the endothelial glycocalyx \cite{becker2015degradation, lipowsky2018role}, the glycocalyx of RBCs also plays a crucial role in maintaining their shape and stability, preventing premature removal from the circulatory system \cite{Mockl}. In certain diseases such as sickle cell anemia, a disrupted glycocalyx is associated with altered cell shape and increased fragility of RBCs, resulting in blockages in small blood vessels and decreased oxygen delivery to tissues \cite{elion2004vaso, maciaszek2011microelasticity}. In diabetes, high glucose levels can affect the structure and continuity of the glycocalyx layer in RBCs, leading to adverse effects, including increased RBC rigidity and enhanced adhesion to blood vessel walls \cite{chang2017modeling, agrawal2016assessment, singh2011high}.

Amylase is a digestive enzyme secreted by the pancreas and salivary glands. The main function of amylase in vivo is to hydrolyze the glycosidic bonds in starch molecules to convert complex carbohydrates to simple sugars \cite{akinfemiwa2020amylase, williams2019amylase}. Elevated levels of serum amylase have been observed in several pathological conditions, including pancreatic disease \cite{Panc}, intestinal disease \cite{banks1984identification}, decreased metabolic clearance and macroamylasemia \cite{berk1967macroamylasemia}.  Diabetic ketoacidosis has also been associated with increased levels of amylase \cite{vantyghem1999changes}.  Given that the glycocalyx is composed of various sugars like glucose, galactose, mannose, and others, which are connected to each other in different ways \cite{tarbell2016glycocalyx}, it is plausible that an increased serum level of amylase may affect the integrity and structure of RBCs glycocalyx.  It has  been shown that the degradation of the RBCs glycogalyx due to enzyme activity enhances  RBC-RBC adhesion under static conditions \cite{pot2011proteolytic}, by using rat blood. However a direct analysis of the role of glycocalyx layer in the aggregation process under flow is still lacking.  

In a completely different context, that of microgravity experienced during long space missions, the metabolic characteristics of astronauts undergo significant changes. Prolonged space flights impact the metabolism, as evidenced by the analysis of capillary blood samples from Russian cosmonauts conducted by Markin et al \cite{markin1998dynamics}. Notably, the study reported an increase in pancreatic amylase enzyme levels compared to basal levels during the space missions.
A more in-depth exploration is essential to understand the pathophysiological processes occurring in spaceflight. This understanding is crucial for guiding risk assessment and developing appropriate preventive, diagnostic, treatment, and countermeasure strategies for cardiovascular complications in space.

The primary objective of our study is to investigate alterations in RBC aggregation following glycocalyx cleavage by amylase, utilizing human blood samples. This investigation was carried out through microfluidic experiments, coupled with an assessment of glycocalyx layer changes using confocal microscopy. Our analysis reveals a significant enhancement in the aggregation process following amylase treatment of RBCs. These findings highlight  the important role of glycocalyx cleavage in shaping the dynamics of RBC aggregation, potentially influencing various physiological processes. Furthermore, numerical simulations have been conducted, and their results align consistently with the observations made in our experimental studies, leading to precious estimate of the correlation between amylase activity and the strength of adhesion energy between RBCs.
\section{Materials and Methods}

\subsection{Red blood cell samples} 
\label{sample}
Whole blood was obtained from the local blood bank ``Etablissement Fran{\c c}ais du Sang'' (EFS, Grenoble). Blood from 5 different single donors was collected in 3mL citrate tubes and stored at 4$^{\circ}$C up to a maximum of 3 days before use. Whole blood was centrifuged at 6000 RPM for 3 minutes, the supernatant was then removed by aspiration and the pelleted Red Blood Cells (RBCs) were resuspended in Phosphate Buffer Saline (PBS 1x, Sigma-Aldrich). Washing of RBCs by centrifugation/resuspension was repeated three times in PBS.

After washing, the glycocalyx of the RBCs was degraded by incubating the cells in PBS solutions of $\alpha$-amylase (type VI-B from porcine pancreas, Sigma-Aldrich) at different concentrations: 200, 500, 1000 and 2000 units/L. $\alpha$-amylase incubation was performed for 3 hours on a tube roller (uniROLLER 6, Lab Unlimited, Ireland) to avoid RBCs sedimentation during treatment. After incubation, RBCs were washed 3 times by centrifugation/resuspension in PBS.

For microfluidic experiments, RBCs were suspended at 5\% hematocrit in PBS containing 15 mg/mL of Dextran 150 ($M_w\simeq 150$ kDa, Sigma-Aldrich) and 1 mg/mL of Bovine Serum Albumin (BSA, Sigma-Aldrich).

For fluorescence microscopy of the glycocalyx, RBCs were stained with fluorescently-labelled Wheat Germ Agglutinin (WGA-Alexa488, Molecular Probe$^{\texttrademark}$) used at 5 $\mu$g/mL concentration in PBS, and subsequently washed  and suspended in PBS at 0.5\% hematocrit before imaging.

\subsection{Microfluidic device}
 \label{microfluidic}

Microchannels  were fabricated using a standard soft-lithography technique. A master mold of the microfluidic device was obtained from SU8 photoresist  spin-coated onto a silicon wafer and exposed to UV light on a scanning laser lithography machine (Dilase 250, KLOE, France). PDMS (Sylgard 184) was cast onto the mold and left for curing for 2 hours at 65$^\circ$C. A glass coverslip (\#1, thickness 150 $\mu$m) was used as the bottom part of the microchip and was
permanently sealed to the PDMS upper part after exposure of the surfaces of both elements to an oxygen plasma. Microfluidic chips used in the present study comprise 10 straight channels in parallel, each of length $L=3$ cm, width $w=45\,\mu$m, and height $h=10\,\mu$m, all connected to a common inlet and outlet. Microfluidic chips were connected to inlet and outlet reservoirs using medical-grade polyethylene tubing of 0.86 mm inner diameter.

A pressure-driven flow was applied in the channels with a mk3 OB-1 pressure controller (Elveflow, France) used to pressurize the inlet reservoir containing the suspension of RBCs. The pressure drop between inlet and outlet was set in the range $\Delta P=30-130$ mbar. Knowing the channel geometry, the pressure drop is converted into an average hydrodynamic shear stress at the wall, $\tau$, by writing the balance of pressure ($\Delta P\times \text{channel cross-section}$) and shear forces ($\tau \times \text{channel inner surface}$) acting on the fluid:
\begin{equation}\label{eq:stress}
\tau=\frac{hw}{2(h+w)}\frac{\Delta P}{L}
\end{equation}
$\Delta P$ in the range $30-130$ mbar thus corresponds to an average wall shear stress $\tau=0.4-1.8$ Pa, encompassing physiological values of the blood circulation.

\subsection{Microscopy}

WGA-stained RBCs were left to sediment on a BSA-coated glass coverslip and imaged by confocal fluorescence microscopy, on a Zeiss inverted microscope equipped with a LSM710 module and using a 40x/NA1.3 oil-immersion objective. Images were taken in raster mode with a lateral size of 512$\times$512 pixels, lateral resolution of 0.263 $\mu$m/pixel, 1.5 $\mu$s pixel dwell time and 8bit depth.

Unstained RBCs, under quiescent or flow conditions, were imaged in bright field on an Olympus inverted microscope, using LED illumination at 455 nm wavelength, a 60x air objective, and a Photron Mini UX50 camera. 12bit monochrome images of 1280$\times$1024 pixels (0.167 $\mu$m/pixel) were acquired at 2000 frames per second under flow conditions.

\subsection{Image processing and analysis}
\label{image processing}

Experiments were performed, for each amylase and flow condition, with blood samples from the 5 individual donors. Collected data were subsequently pooled together for analysis.

The state of the glycocalyx after exposure to amylase was assessed from confocal images of WGA-stained RBCs. For each amylase concentration, the average and standard deviation of the fluorescence intensity was computed over a total of 425 images (85 per donor). The average intensity values obtained for the various amylase concentration were then normalized by that obtained without amylase treatment. 

For RBCs aggregation, bright field image analysis and extraction of relevant information were performed using a custom Python program developed in-house. The program used the 'scikit-image' (skimage) library, which provides a comprehensive set of image processing and analysis tools. For static conditions, a total of 100 images per group of amylase were processed, capturing multiple representative views of the samples. Under flow conditions, an extensive dataset consisting of 20,000 images per group of amylase was processed, accounting for the dynamic nature of the system. First, the program employed a binarization technique to convert the images into binary format, enhancing the contrast between the objects of interest and the background.  Subsequently, object detection algorithms were applied to identify and segment the aggregates of RBCs within the images. The program then labeled each aggregate and calculated the number of RBCs in each aggregate from the extracted aggregate surface (NB: going from area to cell number requires an assumption on the surface of an individual RBC right ? If so, we should state this here). Furthermore, the sphericity of the aggregates was determined by computing the ratio of the aggregate's area and perimeter. The average tube hematocrit was calculated from images as the ratio of the total volume of the tracked RBCs (assumed to be 90 $\mu m^{3}$ per RBC) to the channel volume in the field of view.

\subsection{Numerical simulations} 

 The simulation is performed in 2D by using a lattice Boltzmann method, as described in our earlier studies \cite{kaoui2011two,shen2017interaction}. 
 The fluids inside and outside the cells are described by Navier-Stokes equations (with a negligibly small inertia; see below).
 Several previous studies have supported that most of 2D simulations capture the essential features in 3D \cite{gou2021red, franke2011numerical, abbasi2022dynamics, kaoui2009red,kaoui2011complexity,farutin2014symmetry,abbasi2021erythrocyte,lanotte2016red,mauer2018flow,flormann2017buckling}. 
 We have thus opted for 2D simulation to take benefit from computational efficiency. The membrane force is the bending one supplemented with the constraint of local membrane incompressibility. 
 
For the adhesion between cells, we use a Lennard-Jones potential (following previous studies\cite{brust2014plasma,abbasi2021erythrocyte}) in order to account for aggregation due to dextran molecules. This potential takes the following form :
\begin{equation}
\phi = \varepsilon \left[2\left(\dfrac{h}{r_{ij}}\right)^{6}+\left(\dfrac{h}{r_{ij}}\right)^{12}\right],
\end{equation}
  describing attractive interaction at long ranges and repulsive interaction at short ranges. Here $\mathbf{r}_{ij} = \mathbf{X}_{i}-\mathbf{X}_{j}$, where $\mathbf{X}_{i}$ and $\mathbf{X}_{j}$ are two position vectors of two material points on two different RBCs $i$ and $j$. $h$ is the equilibrium distance between two points of the $i$-th and $j$-th RBC and $\varepsilon$ is the minimum energy associated to this distance.
  In one of our previous studies on RBC-RBC aggregation, we have been  extracted (for healthy RBCs) the strength of adhesion energy between RBCs and its relation to dextran/fibrinogen concentration \cite{brust2014plasma}
  
  Several dimensionless parameters enter our problem:
\begin{itemize}
	\item The Reynolds number:
		
	\begin{equation}
		R_{e}=\dfrac{\rho\dot{\gamma}R^{2}_{0}}{\eta_{ex}},
	\end{equation}
	where $R_0$ is a typical radius of RBC, $\rho$ is the suspending fluid density, $\eta_{ex}$	is its dynamic viscosity and $\dot\gamma$ is the average applied shear rate.
	\item The capillary number, which quantifies the flow strength over bending rigidity of the membrane (in our study, we maintained the capillary number within the range of $30$ to $70$, which corresponds to shear rate values ranging from $550s^{-1}$ to $1000s^{-1}$, which are typical values in microcirculation):
		
	\begin{equation}C_{a}=\dfrac{\eta_{ex}\dot{\gamma}R^{3}_{0}}{k}=\tau_c \dot{\gamma},	\end{equation}
	where $\tau_c={\eta_{ex}R^{3}_{0}}/{k}$ is the shape relaxation time of the vesicle, where $k$ is the bending rigidity of the membrane.		

\item The confinement, which  describes the ratio between effective cell  diameter $2 R_0$ and the channel width $W$:
		
	\begin{equation}
	C_{n}=\frac{2R_{0}}{W}
	\end{equation}
	\item The viscosity contrast, which is the ratio between the viscosities of internal ($\eta_{in}$)  and external ($\eta_{ex}$) fluids:
		
	\begin{equation}
		\lambda=\frac{\eta_{in}}{\eta_{ex}}
	\end{equation}
	\item The reduced area which combine the vesicle perimeter L and its enclosed area A:
		
	\begin{equation}
		\mathcal{A}=\dfrac{4A\pi}{L^{2}}
	\end{equation}
	 $\mathcal{A}=1$ for a circle, and is less than unity for any other shape.  
	\item The dimensionless macroscopic adhesion energy, which  is defined as:
		
	\begin{equation}
	\bar{\varepsilon}_{adh}=\frac{{\varepsilon}R_{0}^{2}}{k}
	\end{equation}
\end{itemize}

  \section{Results and Discussion} 
\subsection{Assessment of glycocalyx changes induced upon amylase treatment}

\begin{figure}[htbp]
  \centering
    \includegraphics[scale=0.3]{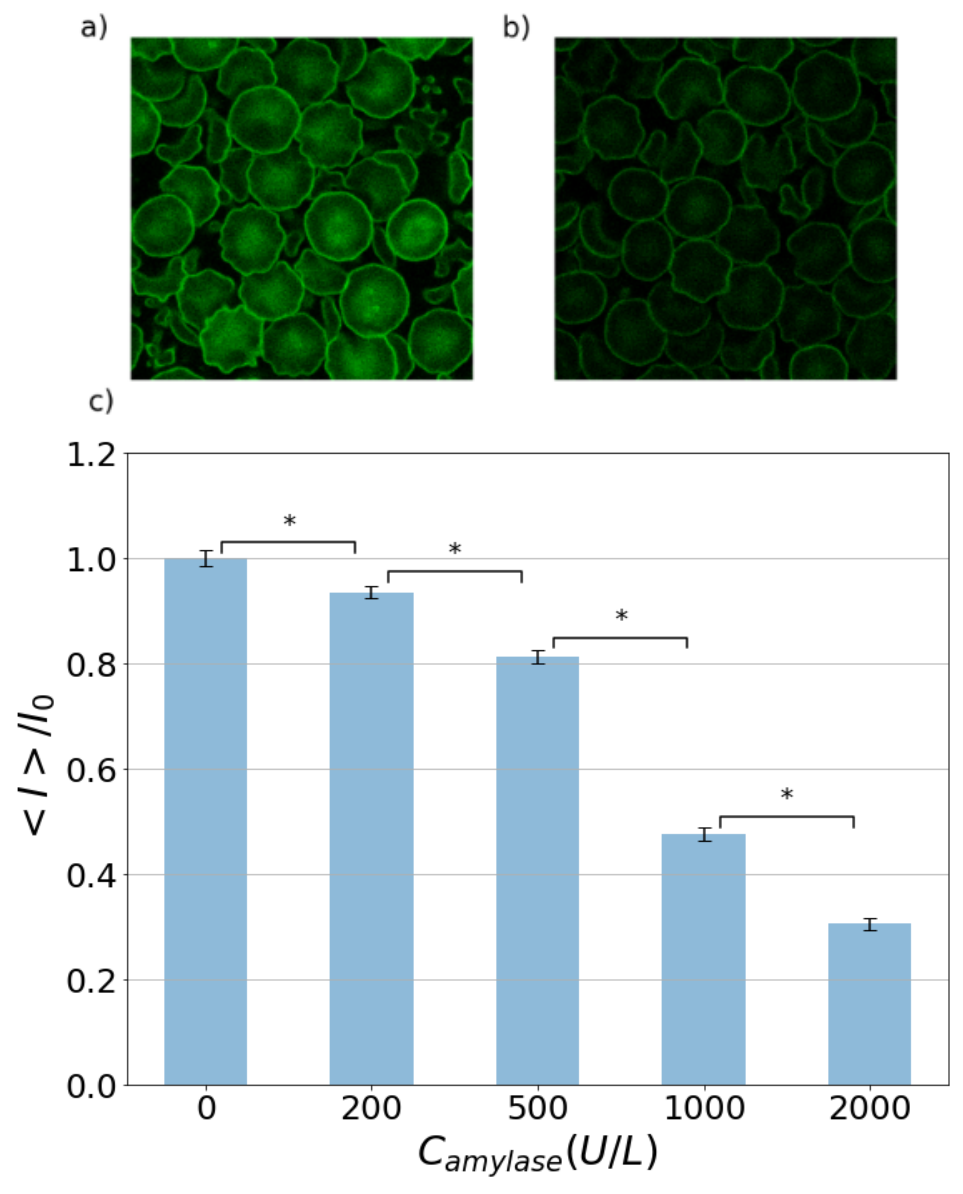}
    \caption{RBCs glycocalyx staining with WGA-Alexa488. a)RBCs untreated with amylase  b) RBCs treated with 2000 U/L amylase. ; c) Normalized fluorescence intensity as a function of amylase concentration $(* : p < 0.05)$ .}
    \label{a}
\end{figure}

Representative confocal images and results of quantitative image analysis are shown in Fig. \ref{a}. We observe that the normalized fluorescence intensity decreases monotonously with the concentration of $\alpha$-amylase to which RBCs were exposed.  This indicates a reduction of the amount of WGA bound to the surface of the cells as they are treated with amylase. WGA is known to bind glycoconjugates (N-acetylglucosamine and sialic acid residues) on cell membranes, and we thus conclude from the observed fluorescence decrease that the glycocalyx of RBCs is degraded by amylase, to a larger extent at higher enzyme concentration. 

The normal range of amylase in human blood is between 70-235 U/L \cite{popper1940pathways,frossard2000early} and can go up to several thousands U/L in pathologies such as pancreatitis \cite{frossard2000early}. The amylase concentrations used in our study thus cover the range from physiological to pathological situations. We note here that under physiological concentration (200 U/L), the fluorescence level is only moderately ($\lesssim$10\%) reduced compared to untreated RBCs, while it drops by more than 50\% for amylase concentrations $\gtrsim$1000 U/L.

\subsection{Aggregation morphology in the absence of flow}
\label{morphology_absence_flow}

To evaluate the aggregability of RBCs  after exposure to different concentration of amylase, the aggregation process was induced by Dextran 150. The concentration of Dextran was fixed at 15 mg/ml for all the experiment in the present article, as the concentration known to induce aggregation similar to that observed in physiological conditions \cite{claveria2016clusters}. The RBCs aggregation is characterized by the number of RBCs per aggregate, $N_{\text{agg}}$, and the aggregate shape parameter (ASP), which is defined as:
\begin{equation} 
ASP = \frac{4 \pi A}{P^{2}} 
\label{ASP-eq} 
\end{equation}
Here, $A$ represents the projected area of the aggregate, and $P$ represents its perimeter. 

\begin{figure}[htbp]
  \centering
    \includegraphics[scale=0.4]{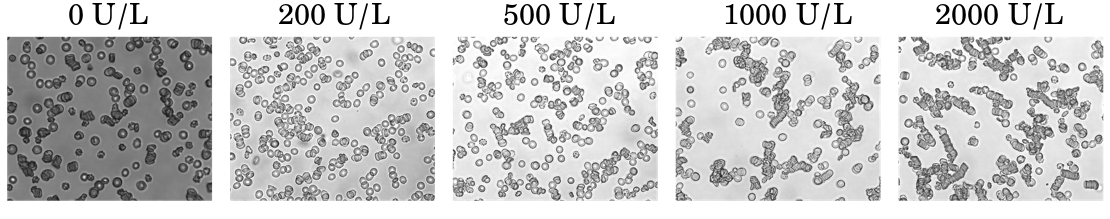}
    \caption{Representative images of RBCs aggregates after treatment with different amylase activity level.}
    \label{b}
\end{figure}

Figure \ref{b}  shows images of typical aggregates obtained for RBCs after treatment with different amylase activities. We qualitatively observe an increase in both the number, the size and the level of branching of aggregates with increasing amylase concentration, particularly in the pathological range of concentrations (500-2000 U/L). 

\begin{figure}[htbp]
  \centering
    \includegraphics[scale=0.2]{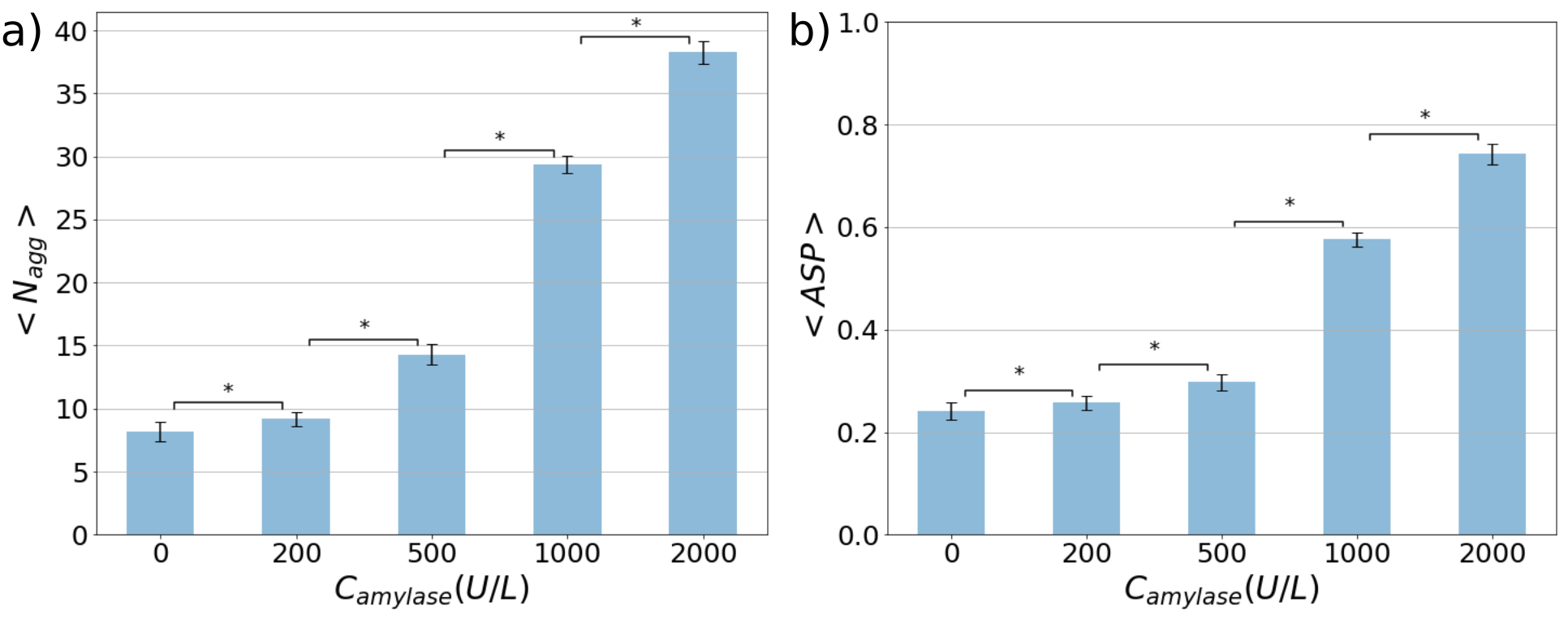}
    \caption{ a) The average number of RBCs per aggregate. b) The asphericity as function of amylase activity level. Data presented as mean$\pm$SEM, $(* : p < 0.05)$.}
    \label{cd}
\end{figure}

We have systematically quantified both the size and morphology of aggregates from 100 images per amylase conditions. Figure \ref{cd}a displays the average number of RBCs per aggregate, $\left< N_{\text{agg}}\right>$, as a function of $\alpha$-amylase activity level. We observe that the average number of RBCs per aggregate remains relatively constant within the physiological range of human blood $\alpha$-amylase \cite{frossard2000early, popper1940pathways}. However, at the activity level of 500 U/L, the average number of RBCs per aggregate shows a slight increase, while higher amylase concentrations ($>$ 1000 U/L) correspond to a significantly higher number of RBCs per aggregate.

The glycocalyx degradation affects not only the number of RBCs per aggregate but also the morphology of aggregates. We use the parameter defined by equation \ref{ASP-eq} to characterize aggregate morphology, where an elongated shape (such as rouleaux shape) corresponds to an ASP value  close to 0.2, whereas a rounded shape has an ASP value close to unity. As shown in Figure \ref{cd}b, for untreated RBCs and at low amylase concentrations ($\lesssim$ 500 U/L), the ASP is smaller than 0.29, corresponding to mostly rouleaux shape of aggregates. However, increasing the amylase activity level (1000 U/L, 2000 U/L) results in much higher ASP values ($>0.5$), indicating a significant deviation from classical rouleaux morphology with branched aggregates of more isotropic shapes. 

\subsection{RBCs aggregates under flow}

\begin{figure}[htbp]
  \centering
    \includegraphics[scale=0.25]{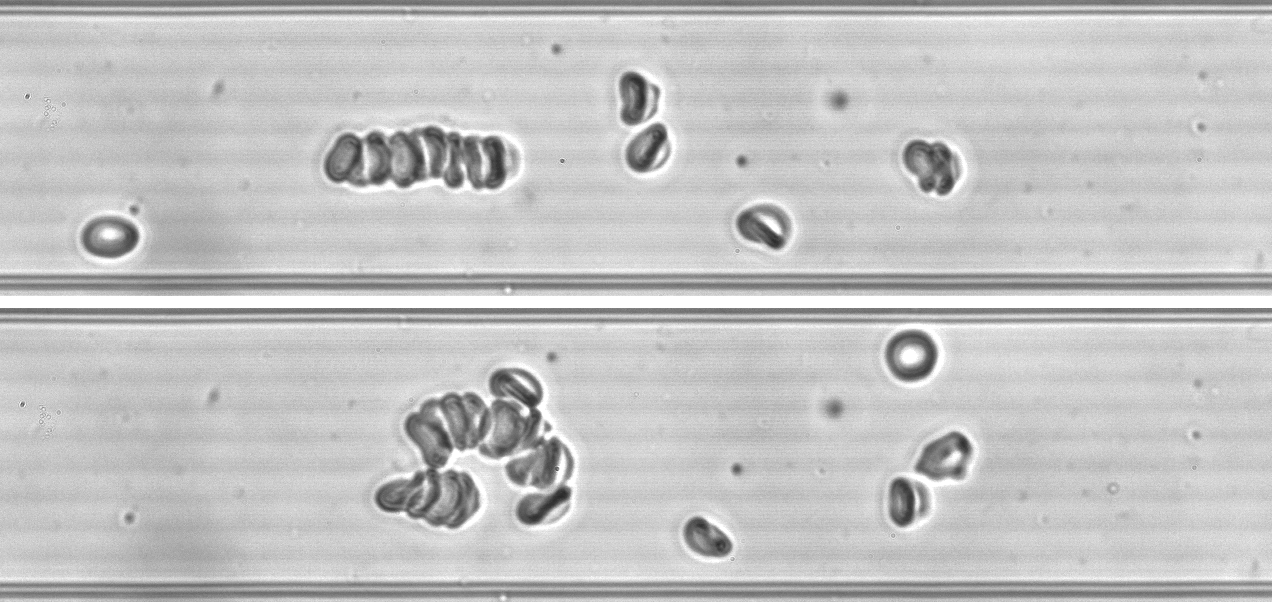}
    \caption{Shapes of aggregates without amylase (top) and with amylase (bottom) with amylase activity level 2000 U/L. Here the wall shear stress is $\tau= 0.4 Pa$}
    \label{f}
\end{figure}

To investigate the impact of RBCs glycocalyx degradation on aggregate stability under flow, microfluidic experiments corresponding to wall shear stress values in the range $0.4 - 1.8$ Pa were conducted for each amylase treatment. Representative images of RBCs in two channels with different amylase treatment (0 U/L and 2000 U/L) are shown in (Fig.\ref{f}), where it can be seen that after amylase treatment the size of the aggregate is larger and is more complex as compared to the classical rouleaux shape. 

\begin{figure}[htbp]
  \centering
    \includegraphics[scale=0.23]{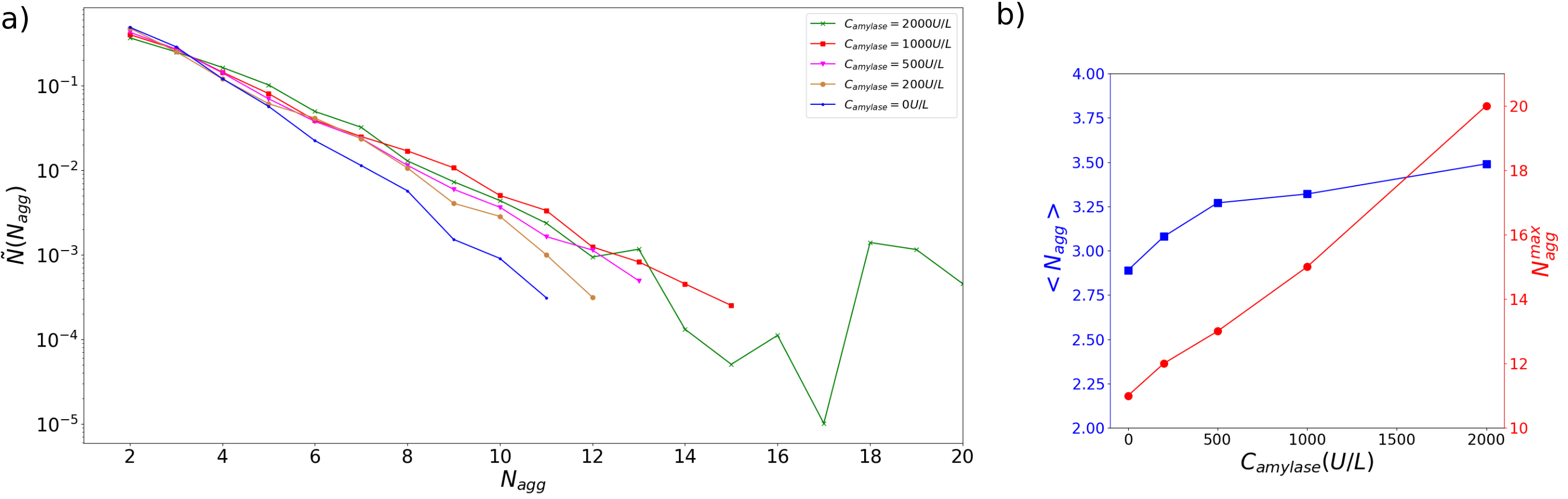}
    \caption{(a) Size distribution of aggregates for each studied $\alpha$-amylase concentration, under a wall shear stress $\tau = 0.4 Pa$. (b) Average ($<N_\text{agg}>$) and maximal ($N^\text{Max}_\text{agg}$) cluster size, computed from the size distributions, as a function of amylase concentration.}
    \label{g}
\end{figure}

The number of aggregates and number of RBCs per aggregate was calculated from 20000 images per $\alpha$-amylase condition. From this, we build the size distribution of aggregates, $\tilde{N}(N_{\text{agg}})$, $i.e.$ the number fraction of aggregates for each observed size. Such distributions are shown in Fig.\ref{g}a, for the various amylase treatments and under a wall shear stress of 0.4 Pa: we observe that the size distributions of aggregates display a very similar quasi-exponential decay, and that the main difference between amylase treatments lies in the cutoff at large sizes. This is recapitulated on Fig. \ref{g}b, where we observe that, under flow, the average number of RBC per aggregate is $\sim 3$ and increases only slightly with amylase concentration (Fig. \ref{g}b), in contrast to quiescent condition. However, the maximum size of the aggregates under flow, $N_\text{agg}^\text{Max}$, is seen to increase markedly with amylase concentration (Fig. \ref{g}b).

\subsection{Evolution of RBCs aggregates under different shear stresses}

\begin{figure}[htbp]
  \centering
    \includegraphics[scale=0.23]{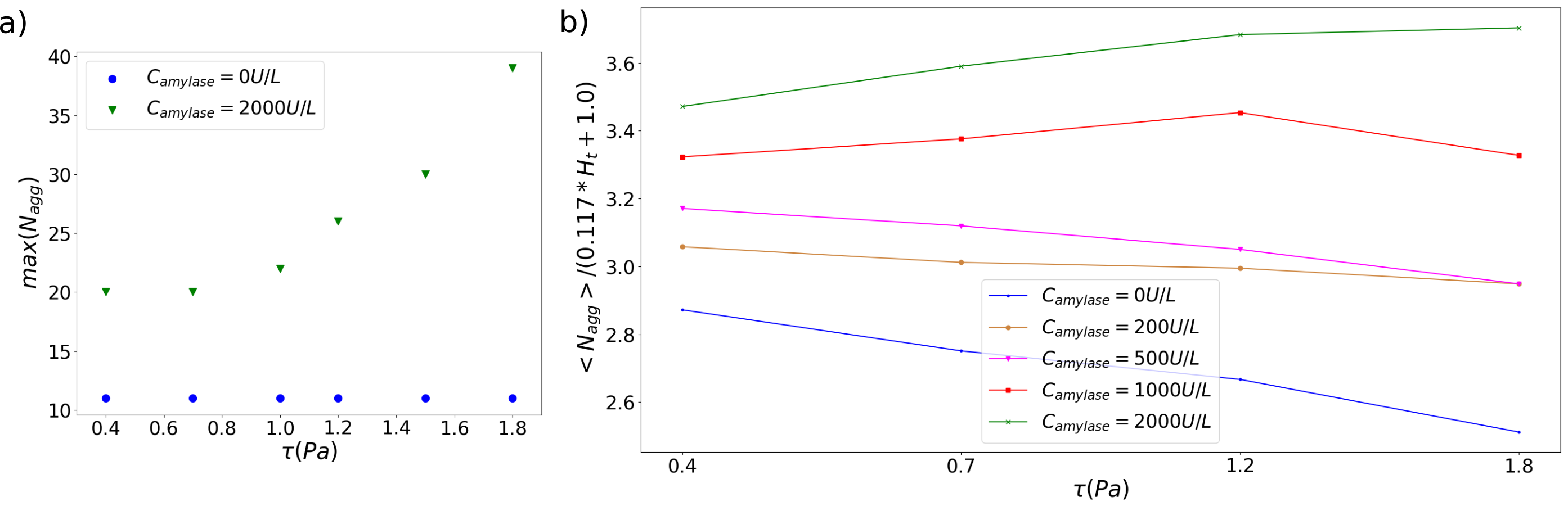}
    \caption{a) The maximum value of the number of red blood cells in aggregates as a function of shear stress at the wall for different value of the amylase concentration. b) Average size of aggregates normalized by a function of hematocrit (dictated by simulations, see below) as a function of amylase activity level.}
    \label{hi}
\end{figure}

We have further investigated the effect of flow strength on the size and shape of aggregates formed under various amylase treatments. From the size distributions of aggregates computed at various flow strengths, we observed that both the average and the maximum size of RBC aggregates would tend to increase under higher shear stresses for amylase-treated RBCs (see $e.g.$ Fig. \ref{hi}a for $N_\text{agg}^\text{Max}(\tau)$ at 2000 U/L amylase concentration.
This is a rather counter-intuitive observation, as one expects higher shear stresses to result in aggregates of smaller sizes because of their fragmentation under stronger flows. However, we have noticed that, upon increasing the flow strength, the tube hematocrit tends to increase in our experiments. Such an increase in local hematocrit is likely to be accompanied by an increased probability of encounter and clustering of aggregates, thus biasing our observations towards larger aggregates under stronger flows. In order to circumvent this uncontrolled change of hematocrit under various flow strength, we rely on numerical simulations to evaluate how hematocrit affects the size of RBCs aggregates (see next section for details). From such numerical simulations, we conclude that, under a given shear stress, the average size of aggregates increases linearly with hematocrit in the range $\text{$H_t$}\simeq5-30$\%. We thus fit our numerical data for $<N_\text{agg}>$ vs $H_t$ by an affine function $f(H_t)$, measure the tube hematocrit in our experiments under various shear strength, as described in section \ref{image processing}, and normalize our experimental data for $<N_\text{agg}>$ by dividing them by $f(H_t)$ in order to compare the various amylase and flow strength conditions under constant hematocrit. 

We show on Fig. \ref{hi}b the evolution of such a normalized  $<N_\text{agg}>$ as a function of shear stress, for the various amylase treatments. We observe that (i) in the range of shear stresses explored, the average size of RBCs aggregates increases upon treatment with higher concentrations of amylase, (ii) while for untreated RBCs, $<N_\text{agg}>$ decreases  with $\tau$, this decrease is less marked for $\alpha$-amylase concentrations of 100 and 500 U/L, and is absent for 1000 and 2000 U/L, for which the average size of aggregates is insensitive to or slightly increases with shear stress.

\begin{figure}[htbp]
  \centering
    \includegraphics[scale=0.19]{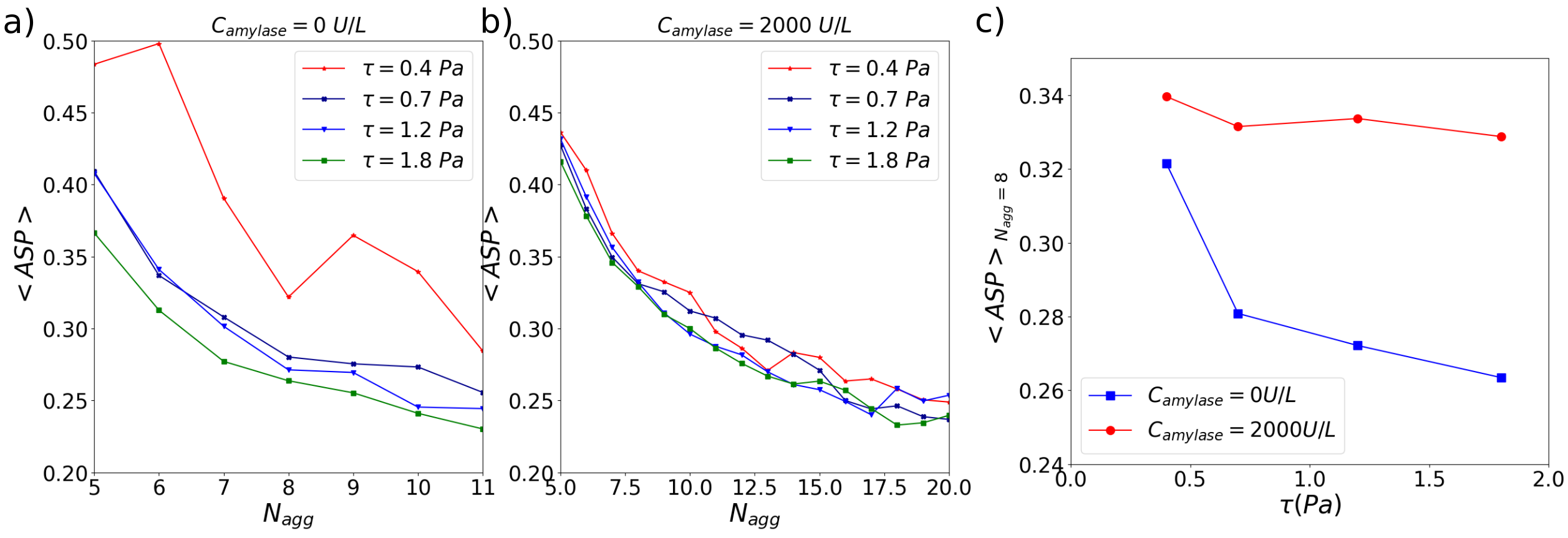}
    \caption{Shape parameter APS as function of cluster size of untreated RBCs agreagates (A) and 2000U/L amylase treated RBCs (B) under the flow with different shear stress at the wall.}
    \label{j}
\end{figure}

The cleavage of glycocalyx by amylase is found to affect also the morphology of aggregates under flow. On Fig. \ref{j}, we plot the Aggregate Shape Parameter (ASP, averaged over aggregates of a given size) as a function of aggregate size, for 4 different shear stresses and 2 types of RBCs (untreated and treated at 2000 U/L amylase). It can be seen, first, that larger aggregates appear to be more elongated and display a lower value of ASP compared to smaller aggregates, and this holds irrespective of the state of the RBC glycocalyx. However, we observe that amylase treatment has an effect on how the ASP is sensitive to the shear stress: for a given size of aggregate, clusters of untreated RBC display an ASP that clearly decreases as the shear stress is increased, indicating less isotropic aggregates at higher flow strength. In contrast to this, RBCs exposed to 2000 U/L of amylase exhibit a much lower sensitivity to shear stress, with an ASP almost independent of flow strength. Such a difference is illustrated on Fig. \ref{j}c for $N_\text{agg}=8$.

\subsection{Numerical simulations}

We conducted simulations with specific dimensionless parameters to capture key aspects of the system. The Reynolds number ($R_e$) was set to $0.1$, ensuring an optimal balance between numerical efficiency and precision for the Stokes regime \cite{takeishi2019haemorheology}. The reduced area ($\mathcal{A}$) and viscosity contrast ($\lambda$) were chosen as $\mathcal{A}=0.64$ and $\lambda=5$, values derived from documented parameters for RBCs. The confinement value ($C_n$) was set to $0.13$, and the capillary number ($Ca$) to $25$, both falling within the experimental range.

\begin{figure}[htbp]
  \centering
    \includegraphics[scale=0.3]{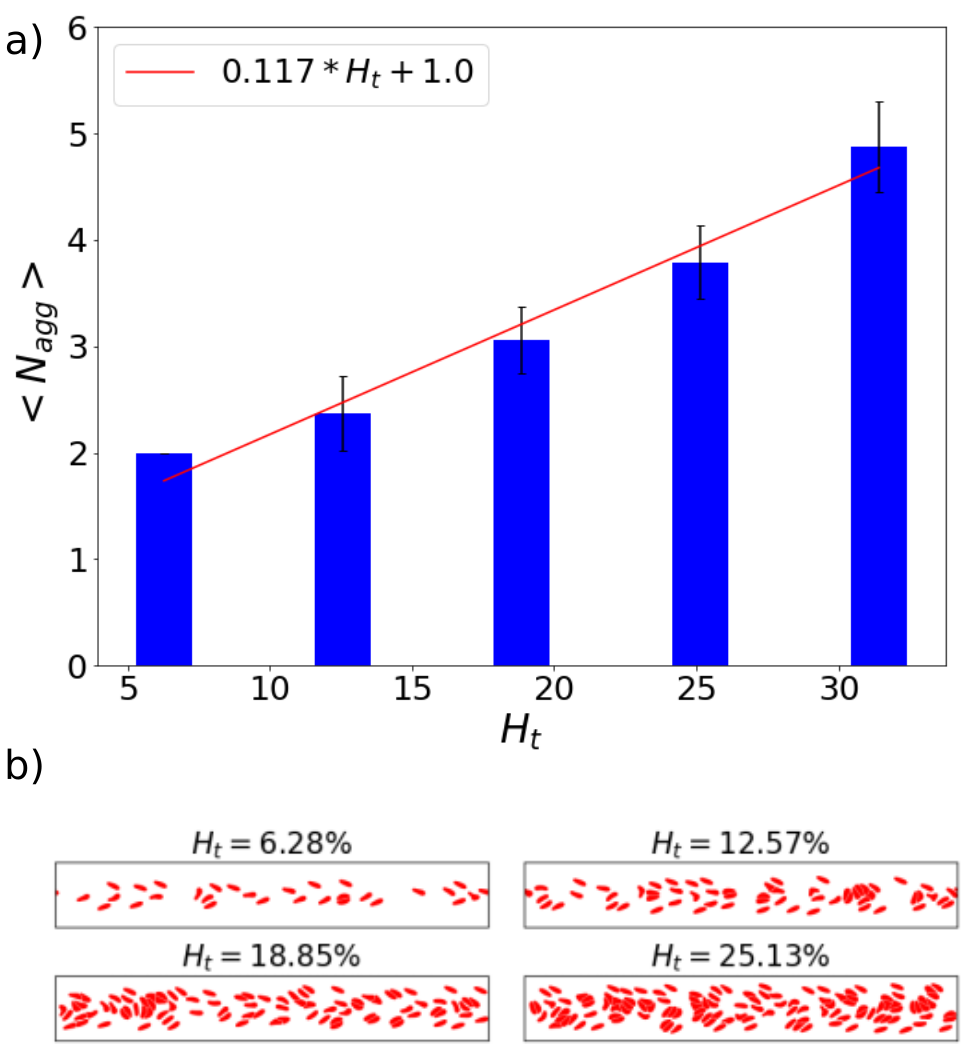}
    \caption{Top: average size of aggregate as a function of hematocrit. Bottom: different suspension configuration}
    \label{k}
\end{figure}

As mentioned above, a first part of our numerical investigation focused on understanding how the size of RBCs aggregates depends on the tube hematocrit. Indeed, we qualitatively expect that an elevation in hematocrit enhances the probability of collision among red blood cells, subsequently increasing both the number and sizes of aggregates.
We have performed a systematic analysis of the statistics of aggregates as a function of hematocrit in our numerical simulations. The results, shown in Figure \ref{k}, unveil a linear increase of aggregate size with hematocrit. These findings are consistent with prior experiments exploring the kinetics of RBC aggregation at various hematocrit \cite{deng1994influence}. The observed increase in size with hematocrit can be fitted empirically by the expression $<N_\text{agg}>=1+0.117\times H_t$.
In an effort to compare our experiments performed at different hematocrits, we thus normalized the experimental data on aggregate size by the previous empirical expression, using for $H_t$ the value measured directly from the obtained images of flowing RBCs. 

\begin{figure}[htbp]
  \centering
    \includegraphics[scale=0.4]{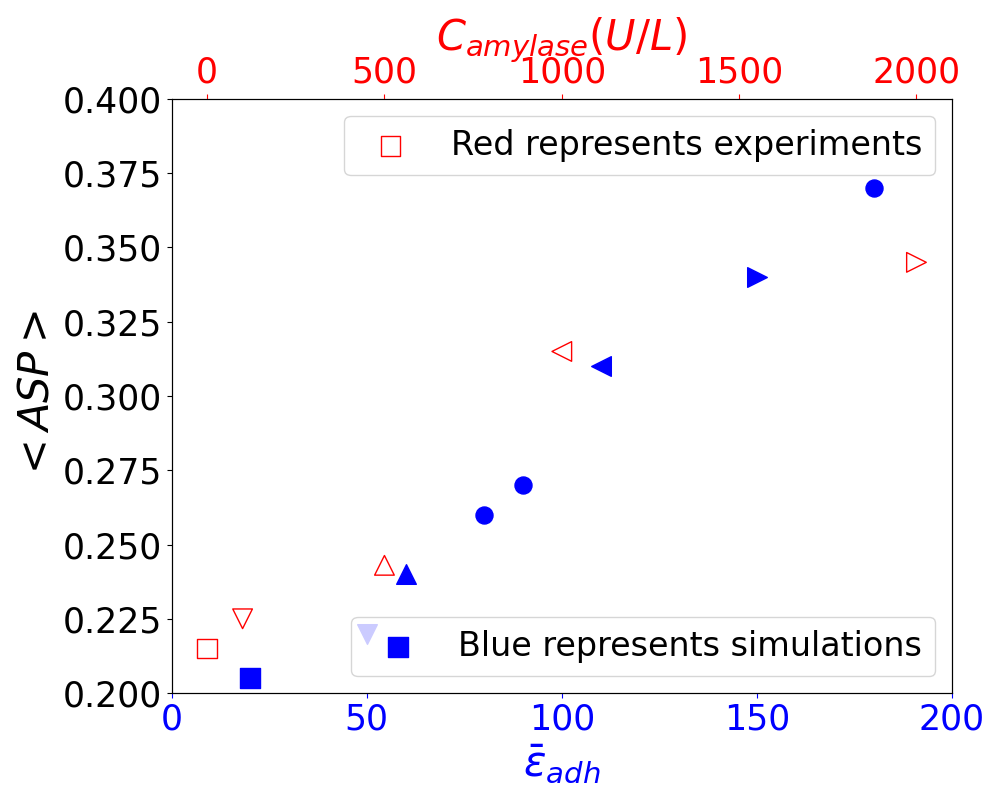}
    \caption{The shape parameters ASP as a function of adhesion energy (lower axis) and corresponding amylase activity (upper axis). We have assigned identical symbols to the experimental data and the best  corresponding result of adhesion energy from the simulation. For example, adhesion energy of 50 (in dimensionless unit) corresponds to an amylase activity of 100 U/L). Filled circles are additional simulation data with no existing experimental data; see text. }
    \label{aspthexp}
\end{figure}

To incorporate the influence of amylase, we adjusted the strength of the potential ($\varepsilon$). For untreated RBCs, the cell-cell interaction potential amplitude was taken from prior Atomic Force Microscopy (AFM) analyses \cite{brust2014plasma}. Comparisons were made between the experimentally obtained shape parameter (ASP) and our simulation results (see Fig. \ref{aspthexp}). For each experimental data (in red in Fig. \ref{aspthexp}) corresponding to different  amylase activities (shown on the upper horizontal axis in Fig. \ref{aspthexp}), we sought the adhesion energy (lower horizontal axis) that best matched the ASP from simulations (by using a hematocrit equal to the average one following from experiments), and have assigned the same symbol (but in blue color) as the symbol corresponding to experimental data.  Physiological values, up to around $\bar \varepsilon_{adh}$ of 80 (as per \cite{brust2014plasma}), align with normal conditions. However, values exceeding this threshold are associated with pathological situations. Notably, amylase activity at approximately 700 U/L and beyond correlates with adhesion energies in the pathological range. This outcome confirms our anticipation that amylase activity in this range significantly impacts RBC-RBC aggregation. Note that data shown with filled blue circles in Fig. \ref{aspthexp} correspond to simulation results having no experimental counterpart (which would have needed several other experiments with different amylase activities to confront with simulations)).

\section{Conclusion}

The present investigation sheds light on the pivotal role played by $\alpha$-amylase in inducing glycocalyx cleavage and subsequently influencing the aggregation of  RBCs. Our study demonstrates that human RBCs exposed to $\alpha$-amylase concentrations ranging from physiological to pathological levels exhibit an increasing trend to (i) form aggregates of larger size, in qualitative agreement with a previous study of the effect of metalloproteinases on rat RBC aggregation \cite{pot2011proteolytic}, and (ii) form clusters of more complex shapes. This holds both under quiescent conditions and under flow. Moreover, we observe that the resistance of aggregates to fragmentation by shear stress is enhanced for amylase-treated RBCs. Finally, our numerical simulations strongly suggest that the effect of exposure to amylase can be accounted for by a change in adhesion energy between RBCs. This points out that glycocalyx degradations induced by $\alpha$-amylase result in a strong change in interaction energy between RBCs. Comparison between experiments and numerical results suggests a more than sixfold increase in adhesion energy when amylase concentration increases from normal to physiopathological level.

The cleavage of glycocalyx thus appears as a significant factor in altering the size, morphology, and stability of RBC clusters. These alterations could potentially give rise to vascular complications, such as  occlusion of microvessels by large, branched and shear-resistant clusters of RBCs, particularly in patients exhibiting elevated levels of amylase activity, as observed in conditions such as pancreatitis. Previous reports have linked pancreatitis to vascular complications, including vessel injury akin to blood wall erosion \cite{barge2012vascular, kirby2008vascular, balachandra2005systematic}.


Furthermore, it is anticipated that amylase activity may not only impact the glycocalyx of RBCs but also extend its influence to that of endothelial cells. This dual effect raises concerns about the potential for endothelial dysfunction, a condition that could facilitate the undesirable adhesion of blood elements to the endothelium. Consequently, a systematic exploration of glycocalyx cleavage in both endothelial cells and RBCs represents a compelling avenue for future research, promising valuable insights into the broader implications of amylase activity on vascular health.


\section*{Conflicts of interest}
The authors declare no competing interests.

\section*{Acknowledgements}

We acknowledge financial support from CNES (Centre National d’Etudes Spatiales), and the French-German University Programme "Living Fluids" (Grant CFDA-Q1-14). The simulations were performed on the Cactus cluster of the CIMENT infrastructure, which is supported by the Rhône-Alpes region (Grant No. CPER0713 CIRA).

\section*{Author contributions statement}

M.A.  conducted the experiments, performed simulations, analyzed and interpreted the data, and wrote the first version of the manuscript. M.J performed systematic experimental analysis.
Y.R performed partial experiments. L.B and D.T interpreted data and revised manuscript. C.M suggested the study, supervised the research, interpreted the data, revised manuscript. All authors reviewed the manuscript. 

\end{document}